\documentclass[aps, prb,superscriptaddress,twocolumn,showpacs,preprintnumbers,amsmath,amssymb]{revtex4}
\usepackage{amssymb}
\usepackage{graphicx}% Include figure files
\usepackage{dcolumn}% Align table columns on decimal point
\usepackage{bm}
\usepackage{subfigure}
\usepackage[normalem]{ulem}

\begin{document}

\preprint{SrZn$_{2}$As$_{2}$/X. J. Yang et al.}
\title{Sr$_{0.9}$K$_{0.1}$Zn$_{1.8}$Mn$_{0.2}$As$_{2}$: a ferromagnetic semiconductor with colossal magnetoresistance}% Force line breaks with \\

\author{Xiaojun Yang}
 \affiliation{Department of Physics and State Key Laboratory of Silicon Materials, Zhejiang University, Hangzhou 310027, China}%Lines break
\author{Qian Chen}
 \affiliation{Department of Physics and State Key Laboratory of Silicon Materials, Zhejiang University, Hangzhou 310027, China}
\author{Yupeng Li}
 \affiliation{Department of Physics and State Key Laboratory of Silicon Materials, Zhejiang University, Hangzhou 310027, China}
\author{Zhen Wang}
 \affiliation{Department of Physics and State Key Laboratory of Silicon Materials, Zhejiang University, Hangzhou 310027, China}
\author{Jinke Bao}
 \affiliation{Department of Physics and State Key Laboratory of Silicon Materials, Zhejiang University, Hangzhou 310027, China}
\author{Yuke Li}
 \affiliation{Department of Physics, Hangzhou Normal University, Hangzhou 310036, China}
\author{Qian Tao}
 \affiliation{Department of Physics and State Key Laboratory of Silicon Materials, Zhejiang University, Hangzhou 310027, China}
\author{Guanghan Cao}
 \affiliation{Department of Physics and State Key Laboratory of Silicon Materials, Zhejiang University, Hangzhou 310027, China}
\author{Zhu-an Xu}
 \email{zhuan@zju.edu.cn}
 \affiliation{Department of Physics and State Key Laboratory of Silicon Materials, Zhejiang University, Hangzhou 310027, China}

\date{\today}% It is always \today, today,

\begin{abstract}
A bulk diluted magnetic semiconductor
(Sr,K)(Zn,Mn)$_{2}$As$_{2}$ was synthesized with decoupled charge
and spin doping. It has a hexagonal CaAl$_{2}$Si$_{2}$-type
structure with the (Zn,Mn)$_{2}$As$_{2}$ layer forming a
honeycomb-like network. Magnetization measurements show that the
sample undergoes a ferromagnetic transition with a Curie
temperature of 12 K and magnetic moment reaches about 1.5
$\mu_{B}$/Mn under $\mu_0H$ = 5 T and $T$ = 2 K. Surprisingly, a colossal negative magnetoresistance,
defined as $[\rho(H)-\rho(0)]/\rho(0)$, up to $-$38\% under a low
field of $\mu_0H$ = 0.1 T and to $-$99.8\% under $\mu_0H$ = 5 T ,
was observed at $T$ = 2 K. The colossal magnetoresistance can be
explained based on the Anderson localization theory.
\end{abstract}


\maketitle

\section{\label{sec:level1}Introduction}

Combining semiconductor and magnetism, diluted magnetic
semiconductors (DMS) which possess both spin and charge degrees of
freedom have attracted much attention because such DMS materials
not only exhibit substantial novel phenomena such as quantum Hall
effects, semiconductor lasers and single-electron charging, but
also bring about numerous applications in sensors, memories as
well as spintronics\cite{1,2,3,4,5}. However, the III-V
based DMS materials, represented by (Ga,Mn)As, obtained by
heterovalent substitution of Mn$^{2+}$ for Ga$^{3+}$, are only
available as thin films due to the limited chemical solubility of
manganese in bulk GaAs ($<$ 1\%), and lack independently controlling of
local moment and carrier densities\cite{14,LiZnMnP}.

The I-II-V based Li(Zn,Mn)As was theoretically proposed to be an n-type DMS and
experimentally synthesized as a bulk p-type DMS material with Curie
temperature (\emph{T}$_{C}$) of 50 K\cite{14,LiZnAs15}. In this
bulk system charge and spin concentrations can be tuned separately
via Li off-stoichiometry and the isovalent substitution of
Mn$^{2+}$ for Zn$^{2+}$, and the solubility of Mn is significantly
enhanced. Actually, only in a limited number of DMS systems the
concentration of acceptor and Mn impurity (i.e., magnetic moment)
can be tuned independently\cite{PbSnMnTe,CdTeMn,ZnMnTe}. Inspired
by the rapid developement of iron-based
superconductors\cite{LaFeAsO,GdTh,Rotter122}, a series of DMS
systems based on the similar layered structure were
found\cite{BaZn2As2,LaZnAsO,LaCuSO,LiZnMnP}.

The so-called colossal magnetoresistance (CMR) is usually observed
in manganites, where the complex and intimate link among magnetic
structure, crystallographic structure and electrical resistivity
makes it a focus of research interest\cite{CMR_LSMO_origin,CMR,
CMR_science}. Materials exhibiting large MR can be exploited to
enlarge the sensitivity of read/write heads of magnetic storage
devices and thus to maximize the information density\cite{CMR}.
But magnetic fields of several tesla are typically required to
obtain such a CMR effect, which limits the potential for
applications\cite{LFMR_APL,LFMR_nature}. The CMR effect under a
low field is highly required from viewpoint of practical use.

SrZn$_{2}$As$_{2}$ is a compound with hexagonal
CaAl$_{2}$Si$_{2}$-type structure (shown in Fig. 1(a) and
(b))\cite{SrZn2As2}, which belongs to P-3m1 (No.164) space group.
The Zn$_{2}$As$_{2}$ layers form a honeycomb-like network. No
detail study on the physical properties of SrZn$_{2}$As$_{2}$ has
been reported yet. Attracting more and more attention, the
honeycomb-like network is essential in recently extensively
investigated topological insulators\cite{Bi2Se3,Bi2Se3_2}.
Recently, a first honeycomb-lattice bulk DMS
(Ba,K)(Cd,Mn)$_2$As$_2$ with $T_C \sim 16$ K\cite{BaCd2As2-D} has been
reported. In this Letter, we report
successful synthesis of a honeycomb-lattice bulk DMS
(Sr,K)(Zn,Mn)$_2$As$_2$ with $T_C \sim $ 12 K and magnetic moment of about 1.5
$\mu_{B}$/Mn under $\mu_0H$ = 5 T and $T$ = 2 K. In this system, charge and spin
degree of freedom can be controlled independently via K$^{+}$ for
Sr$^{2+}$ and isovalent Mn$^{2+}$ for Zn$^{2+}$ substitution,
respectively. Surprisingly, a large MR, defined as
$[\rho(H)-\rho(0)]/\rho(0)$, up to $-$38\% under a low field of
$\mu_0H$ = 0.1 T and to $-$99.8\% under $\mu_0H$ = 5 T at $T$ = 2
K was observed. Only a limited number of systems (systems
represented by (Ga,Mn)As and (La,Sr)MnO$_3$) exhibit such a
remarkable CMR behavior\cite{5,CMR_LSMO_origin,CMR_SSC}. 
The colossal magnetoresistance can be explained based on the Anderson
localization theory. Although $T_C$ of this system may be too low for
practical use, our work encourages people to explore colossal
magnetoresistance in the other recently discovered bulk DMS materials,
such as (Ba,K)(Zn, Mn)$_2$As$_2$\cite{BaZn2As2} and (La,Sr)(Cu,Mn)SO \cite{LaCuSO}
with much high Cure temperatures,
%The colossal
%magnetoresistance can be explained based on the Anderson localization theory.
%No detailed study on magnetoresistance effect was performed on other recently discovered bulk
%DMSs\cite{LiZnAs15,LiZnMnP,BaZn2As2,BaCd2As2-D,LaCuSO,LaZnAsO}. Even
%though the $T_C$ of our system may be too low for practical use,
%the colossal magnetoresistance effect may be found in other recently discovered bulk
%DMSs, in which the $T_C$ were already raised to 180 K for Li(Zn, Mn)As\cite{BaZn2As2}
%and 200 K for (La,Sr)(Cu,Mn)SO, and result in practical use.
which makes the
recently discovered bulk DMSs truly appealing class of systems.

\section{\label{sec:level2}Experimental}

The polycrystalline samples of (Sr,K)(Zn,Mn)$_2$As$_2$ were
synthesized by solid state reaction method. All the starting
materials, Sr granules, K lumps, and the powders of Zn, Mn and As
are of high purity ($\geq$ 99.9\%). As a first step, SrAs was
presynthesized by reacting stoichiometric Sr granules and As
powder at 1123 K for 48 h, and KAs was presynthesized by heating
stoichiometric K lumps and As powder with a ramping rate of 0.25
K/min to 773 K and kept at that temperature for 10 h. Then the
resultants SrAs and KAs, and the powders of Zn, As and Mn were
weighted according to their stoichiometric ratio and then fully
ground in an agate mortar. The mixture was then pressed into
pellets, heated in evacuated quartz tubes at 1223 K for 33 h, and
finally furnace cooled to room temperature. The process was
repeated once again in order to get the pure phase.

Powder x-ray diffraction (XRD) was performed at room temperature
using a PANalytical x-ray diffractometer (Model EMPYREAN) with a
monochromatic CuK$_{\alpha1}$ radiation. The electrical
resistivity was measured by four-terminal method. The dc
magnetization was measured on a Quantum Design magnetic property
measurement system (MPMS-5). The magneto-resistance and Hall
coefficient was measured using a Quantum-Design physical property
measurement system (PPMS).

\section{\label{sec:level3}Results and discussion}

\begin{figure}
\includegraphics[width=8cm]{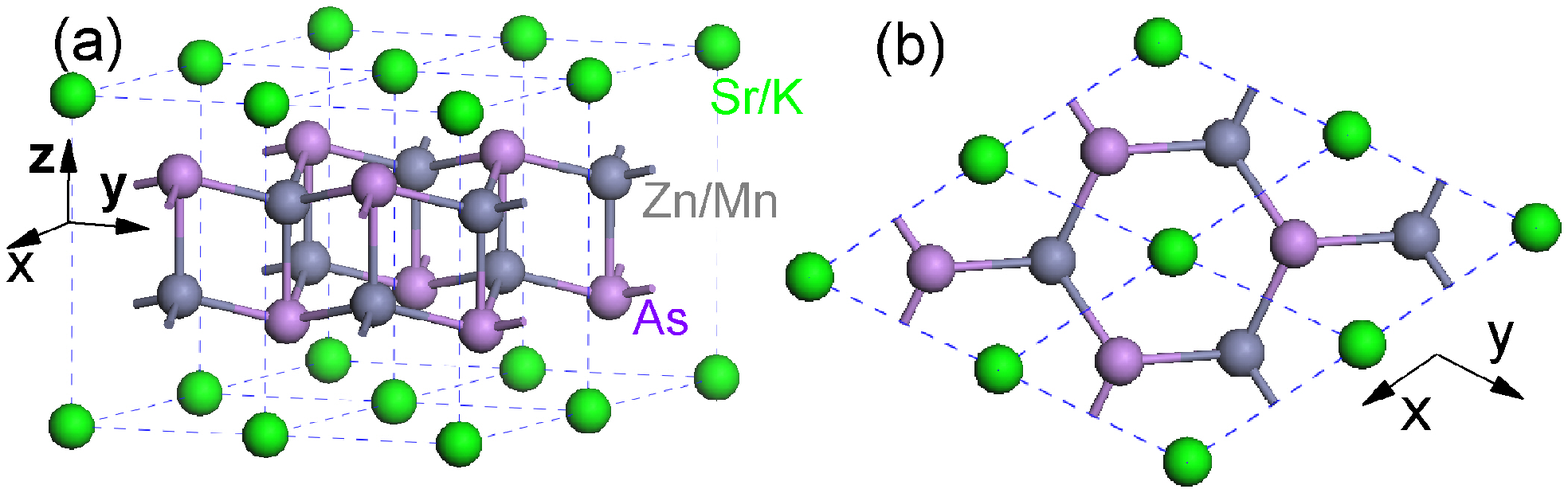}
\includegraphics[width=8cm]{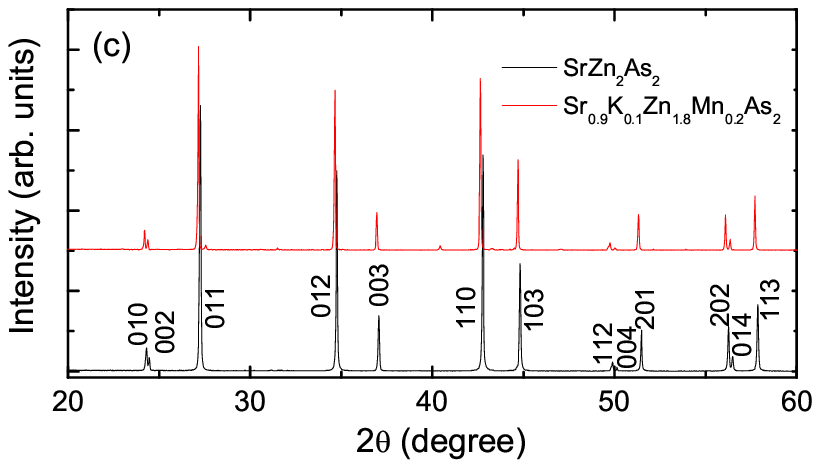}
\includegraphics[width=8cm]{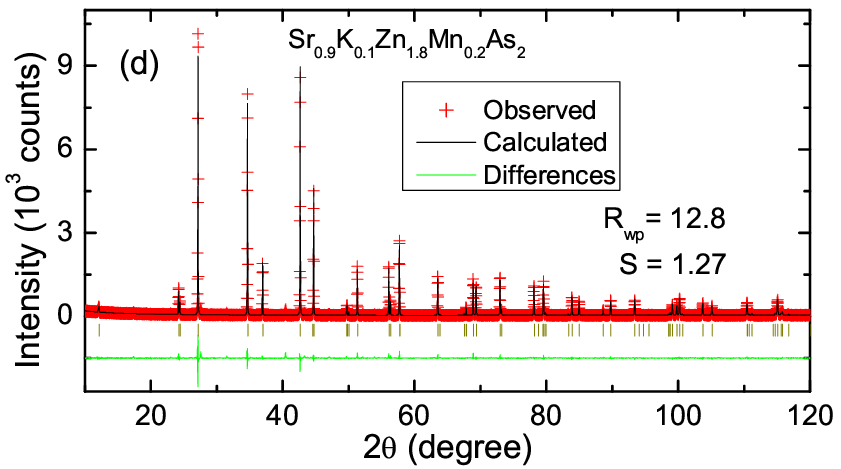}
\caption{\label{Fig.1} (Color online)  (a,b), The crystal
structure of (Sr,K)(Zn,Mn)$_{2}$As$_{2}$. (c), Room-temperature
XRD patterns of SrZn$_{2}$As$_{2}$ (lower curve) and
Sr$_{0.9}$K$_{0.1}$Zn$_{1.8}$Mn$_{0.2}$As$_{2}$ (upper curve),
indexed based on the P-3m1 (No.164) space group. (d), Rietveld
refinement of the powder X-ray diffraction for the
Sr$_{0.9}$K$_{0.1}$Zn$_{1.8}$Mn$_{0.2}$As$_{2}$ sample. }
\end{figure}

Figure 1(c) shows the X-ray diffraction patterns of
(Sr,K)(Zn,Mn)$_{2}$As$_{2}$. The crystal structure is also
sketched in Fig. 1(a) and (b). The diffraction peaks of both
SrZn$_{2}$As$_{2}$ and
Sr$_{0.9}$K$_{0.1}$Zn$_{1.8}$Mn$_{0.2}$As$_{2}$ can be well indexed
based on the P-3m1 (No.164) space group with a hexagonal
CaAl$_{2}$Si$_{2}$-type structure. The XRD patterns indicate that
the samples are essentially single phase. The lattice parameters
of the two samples were obtained by least-squares fit of more than
20 reflection peaks with the correction of zero shift, using space
group of P-3m1 (No.164). The resulting room temperature lattice
constants $a$ and $c$ are 4.223 {\AA} and 7.270 {\AA},
respectively for the SrZn$_{2}$As$_{2}$ parent compound,
consistent with the previously reported values $a$ = 4.223 {\AA}
and $c$ = 7.268 {\AA}\cite{SrZn2As2}. Upon K and Mn doping, the
lattice constants of
Sr$_{0.9}$K$_{0.1}$Zn$_{1.8}$Mn$_{0.2}$As$_{2}$ increase slightly
to $a$ = 4.232 {\AA} and $c$ = 7.280 {\AA}, which is consistent
with the fact that the ionic radius of the Mn$^{2+}$ ion is larger
than that of Zn$^{2+}$ and the ionic radius of the K$^{+}$ ion is
also larger than that of Sr$^{2+}$. As shown in Fig. 1(d), the
Rietveld refinement of the
Sr$_{0.9}$K$_{0.1}$Zn$_{1.8}$Mn$_{0.2}$As$_{2}$ sample based on
the CaAl$_{2}$Si$_{2}$-type structure shows that the calculated
profile well matches the experimental data. The weighted reliable
factor $R_{wp}$ and the goodness of fit $S$ are 12.8\% and 1.27,
respectively.

\begin{figure}
\includegraphics[width = 8cm]{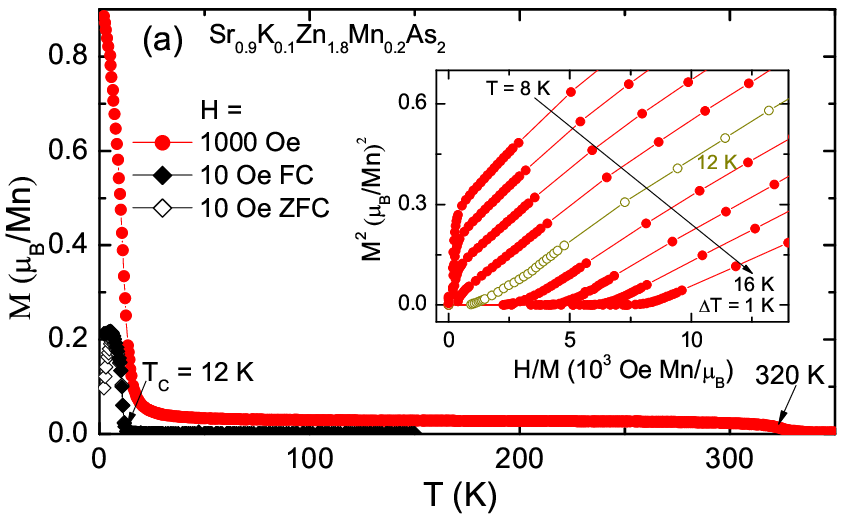}
\includegraphics[width = 8cm]{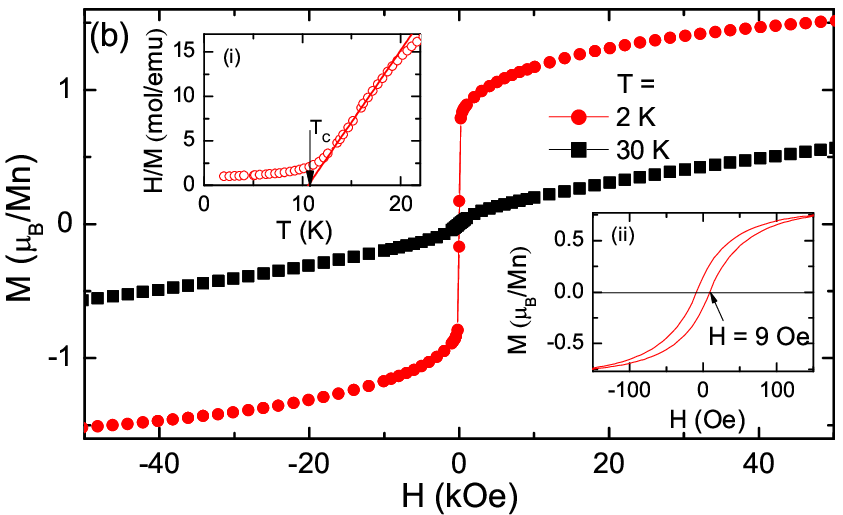}
\caption{\label{Fig.2}(Color online) (a), Temperature dependence of dc
magnetic susceptibility measured under \emph{H} = 1 kOe  and 10 Oe for
Sr$_{0.9}$K$_{0.1}$Zn$_{1.8}$Mn$_{0.2}$As$_{2}$ with solid symbols standing for FC and open
ones for ZFC. No significant difference can be found between ZFC and
FC data when \emph{H} = 1 kOe. The inset of (a) display plot of
$M^2$ verse $H/M$ (Arrott plot) at various temperatures. (b), Field dependence of
magnetization measured at 2 K and 30 K. The inset:of (b): (i) plot of $H/M$
vs. $T$ and the determination of the Curie temperature $T_C$ is shown; (ii)
the enlarged $M(H)$ curve.}
\end{figure}

In Fig. 2(a), the temperature dependence of dc magnetization
of Sr$_{0.9}$K$_{0.1}$Zn$_{1.8}$Mn$_{0.2}$As$_{2}$
is shown for both zero-field cooling (ZFC) and field-cooling (FC)
procedures under \emph{H} = 1 kOe  and 10 Oe. No significant
difference can be found between ZFC and FC data when the applied
field \emph{H} = 1 kOe. Below $(T_{C}) \sim $ 12 K, an abrupt
increase in magnetization can be observed, which is a
signature of ferromagnetic order. In the inset of Fig. 2(a), the
data are shown in the plot of $M^2$ verse $H/M$ (Arrot
plot), which clearly indicates that the Curie temperature ($T_C$)
is about 12 K. Under an applied field of 10 Oe, a bifurcation
between ZFC and FC curves can be observed below the temperature
$T_f$ = 6 K, where $T_f$ stands for the freezing temperature of
individual spins or domain wall motion. As discussed in Ref.
\cite{LaZnAsO, Sr32522}, the bifurcation of ZFC and FC curves and
the hysteresis loops can be found not only in regular ferromagnets
but also in spin glasses. In typical spin glass systems, the
moment size is usually small, i.e., $\sim$ 0.01 $\mu_B$/Mn for
the II-VI (Zn,Mn)Se or other typical diluted alloy spin
glasses\cite{Tholence,Monod,Prejean}. Considering that the
magnetic moment of our system is 0.22 $\mu_B$/Mn under a
small field of $H$ = 10 Oe and can reach 0.89 $\mu_B$/Mn under $H$
= 1000 Oe, we tentatively assign it to a ferromagnetic ordering
rather than a spin glass. The small magnetization anomaly at $T$
$\sim$ 320 K should be due to the traceable MnAs impurity
phase,\cite{MnAs,MnAs2} whose content is too small to be detected
by X-ray diffraction. The large increase of magnetization and
magnetoresistance (MR) below temperature of about 12 K cannot be
attributed to the traceable MnAs impurity phase. %However, due to
%the MnAs impurity phase, the $M(T)$ data can not be well described
%by the Curie-Weiss law, and the plot of 1/$(\chi-\chi_0)$ vs. $T$
%shows a linear behavior only in a short temperature range above
%$(T_{C}) \sim $ 12 K (inset (i) of Fig. 2(b)), the obtained
%effective paramagnetic moment value about 8 $\mu_B$/Mn was larger
%than 5.9 $\mu_B$/Mn (considering $S$ = 5/2).
In the inset (i) of Fig. 2(b), we plot $H/M$ vs. $T$ curve,
which goes to zero at $T_C$ (around 12 K), consistent with the Arrot plot

The magnetic field dependence of magnetization, i.e.  $M(H)$
curves, of Sr$_{0.9}$K$_{0.1}$Zn$_{1.8}$Mn$_{0.2}$As$_{2}$ at $T$
= 2 K and 30 K is shown in Fig. 2(b). At $T$ = 2 K, the magnetization
reaches 1.5 $\mu_{B}$/Mn under an applied field of $H$ = 50 kOe,
which is comparable with the result in Li(Mn, Zn)As\cite{LiZnAs15} and (Ba, K)(Zn, Mn)$_2$As$_2$\cite{BaZn2As2}.
At $T$ = 30 K, the $M(H)$ curve indicates that the sample is mainly in a paramagnetic state,
but the traceable MnAs impurity with FM order can be observed.  As shown in
the inset (ii) of Fig. 2(b), the coercive field of
Sr$_{0.9}$K$_{0.1}$Zn$_{1.8}$Mn$_{0.2}$As$_{2}$ is less than 10 Oe
at $T$ = 2 K . Such a coercive field is even smaller than that of
Li(Mn,Zn)As (30-100 Oe)\cite{LiZnAs15}, thus the material should
be appealing in spin manipulation.

\begin{figure}
\includegraphics[width=8cm]{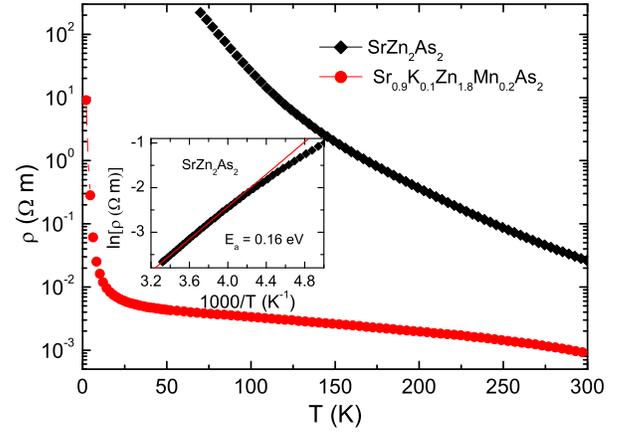}
\caption{\label{Fig.3}(Color online)  Temperature dependence of
resistivity of SrZn$_{2}$As$_{2}$ and
Sr$_{0.9}$K$_{0.1}$Zn$_{1.8}$Mn$_{0.2}$As$_{2}$. The inset: the
 ln$\rho$ vs. $1/T$ plot for SrZn$_{2}$As$_{2}$.}
\end{figure}

As shown in Fig. 3, for the SrZn$_{2}$As$_{2}$ parent compound,
its resistivity clearly exhibits thermally activated behavior with
decreasing temperature, and it increases beyond our measurement
limitation below 70 K. The thermal activation energy ($E_{a}$)
obtained by fitting with the thermal activation formula $\rho(T )
= \rho_{0}$ exp$(E_{a}/k_{B}T$) for the temperature range from 250
to 300 K is 0.16 eV, as shown in the inset of Fig. 3. With 10\%
K-for-Sr substitution, p-type carriers can be introduced, and the
resistivity of Sr$_{0.9}$K$_{0.1}$Zn$_{1.8}$Mn$_{0.2}$As$_{2}$
decreases rapidly compared with the SrZn$_{2}$As$_{2}$ parent
compound, but it still increases with decreasing $T$, which should
be due to the localization effect.
The Hall effect of Sr$_{0.9}$K$_{0.1}$Zn$_{1.8}$Mn$_{0.2}$As$_{2}$ was also
measured at $T$ = 300 K and the charge carrier density is
calculated to be 7.6 $\times$ 10$^{18}$/cm$^{3}$ by using the
one-band model formula $R_{H}$ = $1/ne$. The positive Hall
coefficient demonstrates that the hole type charge carrier is
indeed dominant in the system, consistent with that 10\% K-for-Sr
substitution. However, the Hall coefficient is hard to measure at low
temperatures, due to the extremely large resistivity.

%Sr$_{0.9}$K$_{0.1}$Zn$_{1.8}$Mn$_{0.2}$As$_{2}$ becomes
%ferromagnetic below $T_C$ of 12 K, but it still shows
%semiconducting behavior. This feature implies that a sizable
%exchange interaction between Mn moments can be mediated by charge
%carriers (holes) before they become fully itinerant, and that the
%existence of the metallic state is not a precondition for
%formation of a homogeneous ferromagnetic state\cite{SR}.
%Ferromagnetism in insulating materials could involve the
%hybridization of locally polarized valence band states and Mn
%impurity states where the Fermi level lies between the impurity
%bound states and the valence band\cite{SR,JPSJ_MI,PRB_MI}.

\begin{figure}
\includegraphics[width = 8cm]{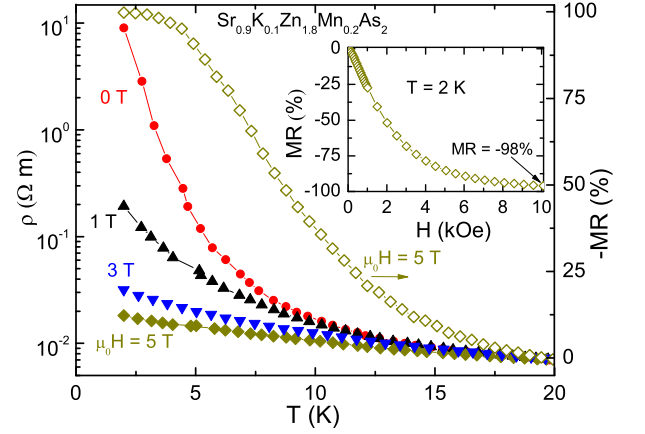}
\caption{\label{Fig.4}(Color online) Temperature dependence of
resistivity (left axis) under $\mu_{0}H$ = 0, 1, 3, 5 T and
magnetoresistance (MR) (right axis), defined as
$[\rho(H)-\rho(0)]/\rho(0)$,
 under $\mu_{0}H$ = 5 T for Sr$_{0.9}$K$_{0.1}$Zn$_{1.8}$Mn$_{0.2}$As$_{2}$. The inset displays the field
 dependence of MR at $T$ = 2 K.}
\end{figure}

We then measured the magnetoresistance(MR), defined in the standard way as
$[\rho(H)-\rho(0)]/\rho(0), $ (The magnitude of MR could be much larger if
it is defined in the other way as $[\rho(H)-\rho(0)]/\rho(H)$), for
Sr$_{0.9}$K$_{0.1}$Zn$_{1.8}$Mn$_{0.2}$As$_{2}$, as shown in Fig.
4. Without magnetic field, the resistivity is 9.0 $\Omega$ m at 2
K, which decreases rapidly with applied magnetic field. It reaches
5.84$\Omega$ m under a low field of $\mu_{0}$\emph{H} = 0.1 T (MR
= $-$38\%), then to 0.19 $\Omega$ m under $\mu_{0}$\emph{H} = 1 T
(MR = $-$98\%), and even to 0.018 $\Omega$ m under
$\mu_{0}$\emph{H} = 5 T (MR = $-$99.8\%). Such a colossal
magneto-resistance has only been observed in a limited number of
systems (represented by (Ga,Mn)As and (La,Sr)MnO$_3$)\cite{5,CMR_LSMO_origin,CMR_SSC}. These materials have attracted
much attention in the field of condensed matter physics. %Our
%discovery should bring new excitement to this field and make the
%recently discovered bulk
%DMSs\cite{LiZnAs15,LiZnMnP,BaZn2As2,BaCd2As2-D,LaCuSO,LaZnAsO}
%truly appealing class of systems.

In a typical manganite CMR system, i.e. (La, Sr)MnO$_{3}$, the
scattering by fluctuating local magnetic moments plays an
important role in the charge transport\cite{CMR_LSMO_origin}, and
thus MR is large around $T_{C}$, but it usually becomes smaller far below
$T_{C}$\cite{CMR_LSMO_PRL}. However, CMR in
Sr$_{0.9}$K$_{0.1}$Zn$_{1.8}$Mn$_{0.2}$As$_{2}$ keep increasing
below $T_C$, which implies that the mechanism of CMR in this
system could essentially be different. CMR in this system have two
interesting features: (1) low coercivity field, (2) maximum in MR
at temperatures far below $T_C$.

A similar negative magnetoresistance was observed in
p-(Zn,Mn)Te, which also occurred below $T_C$ and was thought to be
related with the localization effect\cite{p_ZnMnTe}. On crossing
the metal-insulator transition (MIT), the extended states become
localized. However, according to the scaling theory of the MIT,
their localization radius $\xi$ decreases rather gradually from
infinity at the MIT toward the Bohr radius deep in the insulator
phase, so that on a length scale smaller than $\xi$, the wave
function retains an extended character\cite{science-zener}.
Previous studies have demonstrated that ferromagnetic interactions
can be mediated by the weakly localized holes showing an extended
character\cite{science-zener,MIT_JPSJ}.

\begin{figure}
\includegraphics[width = 8cm]{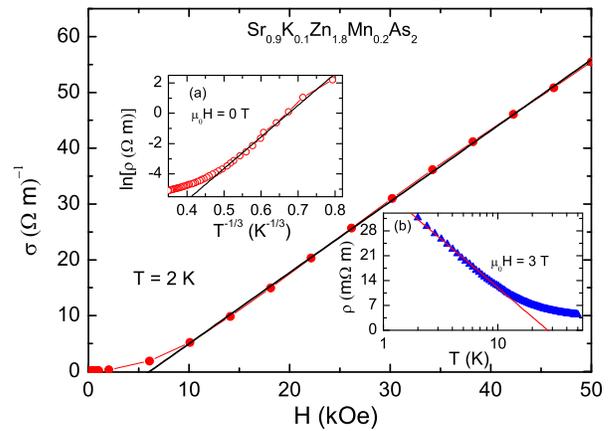}
\caption{\label{Fig.5}(Color online) Magnetic field dependence of
conductivity of Sr$_{0.9}$K$_{0.1}$Zn$_{1.8}$Mn$_{0.2}$As$_{2}$ at
$T$ = 2 K. Inset (a): Resistivity of
Sr$_{0.9}$K$_{0.1}$Zn$_{1.8}$Mn$_{0.2}$As$_{2}$ under zero
magnetic field in $ln\rho$ vs. $T^{-1/3}$  (2D-VRH) plot; (b)
resistivity of under $\mu_0H$ = 3 T in $\rho$ vs. $logT$ plot.}
\end{figure}

In an Anderson localized system, the Fermi level is located on the
localized side of the mobility edge. The application of a magnetic
field introduces Zeeman shifts of each eigenstate dependent on
spin directions, and the repopulation among Anderson localized
states. For one of the spin subbands, the mobility edge moves
toward the Fermi level. This results in a negative
magnetoresistance because of the exponential dependence of the
wave function ($\psi(r) \propto e^{-r/\xi}$) overlap on the
localization length $\xi \propto (E_C-E)^\nu$ , where $E_C$ is the
mobility edge and $E$ is the energy of the localized state, $\nu$
is an exponent on the order of unity\cite{MR_ALS, CMR_SSC}. We
plot in Fig. 5 the magnetic field dependence of conductivity
($\sigma = 1/\rho$), and find a remarkable result that $\sigma$ is
linear in $H$ for 10 kOe $\leq H \leq$ 50 kOe, ie. $\sigma$ =
$C(H-H_c)$, with $C$ = 1.27 ($\Omega$ m kOe)$^{-1}$ and $H_c$ =
6.08 kOe. Based strictly on the concepts of localization, Abrahams
$et$ $al.$\cite{Local_T1} and later Imry\cite{Local_T2} pointed out,
using scaling arguments, that $\sigma \propto (E_{F}(0) -
E_{C})^{\nu}$, where $\nu$ = 1, which was also observed experimentally
in several magnetic systems\cite{CMR_PRL,upturn-1_PRL}. Furthermore, it is known that, on
the insulating side, $|E_F(0) - E_C|$ is linear in $H$ for fields
beyond approximately 6 kOe\cite{CMR_PRL,E_pro_H}. We argue that
$E_F(0) - E_C \geq $ 0 in our case, and then $\sigma
\propto (E_F(0) - E_C)^{\nu} \equiv C(H-H_c)^{\nu}$. Therefore,
$\nu$ = 1 observed in the present MR data may be compared directly
to the exponents derived from the scaling theories\cite{CMR_PRL}.

Exhibiting rapid increase at low temperatures, the resistivity can
be roughly fit with $\rho (T) = Ce^{(T_0/T)^{1/3}}$ under $\mu_0H$ =
0 T (shown in the inset (a) of Fig. 5), which should be a typical
behavior of Anderson insulators and known as the two-dimensional
variable range hopping (2D-VRH)\cite{VRH}, stemming from the
random potential scattering contributed by large numbers of
disorders or defects. Magnetic field can align the magnetic
moments and reduce the disorder of the system, and then
resistivity exhibits the $\rho \propto -logT$ behavior when the
applied field $\mu_0H \geq $ 3 T (shown in the inset (b) of Fig.
5), which might be attributed to quantum correlations to the
conductivity in weakly localized
regime\cite{MIT_JPSJ,upturn-1_PRL,upturn-2_PE}. All the results
indicate that our studied system can be well described by the
localization theories, and the negative CMR could be understood
based on the Anderson localization.

%For a magnetic semiconductor (DMS), there could exists another
%mechanism of MR under high magnetic field. The charge carrier and
%the surrounding cloud of Mn$^{2+}$ spins is likely to form a bound
%magnetic polaron via the p-d exchange interaction.
%\cite{CMR_MP_book} With an external magnetic field, the Mn$^{2+}$
%spins will be in a more ordered state, the magnetic polarons move
%through a sea of Mn$^{2+}$ spins more easily. The negative
%magnetoresistance is attributed to smearing of the polaronic cloud
%with increasing magnetic field\cite{CMR_SSC,CMR_PRL}. Actually,
%the slowly saturating magnetic moment in the magnetization curve
%even at $T$ = 2 K (see the inset of Fig.2) implies the residual
%paramagnetic component of magnetic moments even in the FM state
%\cite{CMR_SSC}.

Since most applications of MR effect require operating magnetic
fields of less than several hundreds of Oe, to reduce the required
magnetic field for CMR in manganites has been a major
goal\cite{LFMR_APL,LFMR_nature}. In the studied
Sr$_{0.9}$K$_{0.1}$Zn$_{1.8}$Mn$_{0.2}$As$_{2}$ system, MR reaches
-38\% only under an applied field of 1000 Oe. The low field MR
%Such low-field CMR makes it an appealing system.
may stem from the spin-dependent scattering at domain walls or
grain boundaries in our poly-crystalline
sample\cite{LFMR_APL,LFMR_nature}. An external field can align the
magnetic domains, and thus the spin-dependent scattering will be
reduced, resulting in a negative MR. Meanwhile, under high
magnetic field, based on the the Anderson localization theory, the
high magnetic field causes a delocalization effect, and thus a
negative colossal magnetoresistance is induced.

It should also be mentioned that the CMR effect is often
accompanied by a metal-insulator phase transition induced by
applying magnetic field near magnetic order temperatures
as observed in III-V based (Ga,Mn)As and (In,Mn)As\cite{SR,WL_JMMM,WL_PRL}. However, no such
metal-insulator phase transition is observed in
Sr$_{0.9}$K$_{0.1}$Zn$_{1.8}$Mn$_{0.2}$As$_{2}$ or a similar
system (Ba, K)(Cd, Mn)$_{2}$As$_{2}$ \cite{BaCd2As2-D}. It is
interesting whether a metal-insulator phase transition could be
induced by applying an even higher magnetic field in these
systems.

\section{\label{sec:level4}Concluding remarks}

In summary, we have successfully prepared a bulk hexagonal DMS,
(Sr,K)(Zn,Mn)$_{2}$As$_{2}$, with \emph{T}$_{C}$ of 12 K, and  
magnetic moment reaches about 1.5 $\mu_B$/Mn under
$\mu_0H$ = 5 T and $T$ = 2 K. It is a soft magnet with a
relatively small coercive field of less than 10 Oe. A low-field
CMR, defined as $[\rho(H)-\rho(0)]/\rho(0)$,
reaches $-$38\% under $\mu_0H$ = 0.1 T and
up to $-$99.8\% under $\mu_0H$ = 5 T. Further studies
suggest that the Anderson delocalization due to applying magnetic
field may account for the observed CMR effect. With decoupled
charge and spin doping, the hexagonal (Sr,K)(Zn,Mn)$_2$As$_2$
combines semiconductor, ferromagnetism and colossal
magnetoresistance, which makes it an appealing system.

\acknowledgments
This work is supported  by the National Basic Research Program of
China (Grant Nos. 2011CBA00103 and 2012CB821404), NSF of China
(Contract Nos. 11174247 and U1332209), Specialized Research Fund
for the Doctoral Program of Higher Education (Grant No.
20100101110004), and the Zhejiang Provincial Natural Science
Foundation of China (Grant. No. Y6100216).

\end{document}